\documentclass[sigconf]{acmart}

\usepackage{booktabs} %

\usepackage[multiple]{footmisc}
\usepackage{hyperref}
\usepackage[most]{tcolorbox}
\usepackage{amsmath}
\usepackage{bm}
\usepackage[caption = false]{subfig}
\usepackage{graphicx}
\usepackage{enumitem}

\usepackage{xcolor}
\usepackage{colortbl}
\usepackage{verbatim}

\usepackage{multirow}

\newcommand{\matr}[1]{\mathbf{#1}}

\newcommand\numberthis{\addtocounter{equation}{1}\tag{\theequation}}

\renewcommand\footnotetextcopyrightpermission[1]{} %
\copyrightyear{2018}
\acmYear{2018}
\setcopyright{rightsretained}
\acmConference{RecSys '18}{}{October 2, 2018, Vancouver, Canada}
\acmDOI{N/A}
\acmISBN{}
\acmPrice{}
\begin{document}

\title{Trust-Based Collaborative Filtering: Tackling the Cold Start Problem Using Regular Equivalence}

\author{Tomislav Duricic}
\affiliation{%
  \institution{Know-Center GmbH}
  \city{Graz} 
  \state{Austria} 
  }
\email{tduricic@know-center.at}

\author{Emanuel Lacic}
\affiliation{%
  \institution{Know-Center GmbH}
  \city{Graz} 
  \state{Austria} 
  }
\email{elacic@know-center.at}

\author{Dominik Kowald}
\affiliation{%
  \institution{Know-Center GmbH}
  \city{Graz} 
  \state{Austria} 
  }
\email{dkowald@know-center.at}

\author{Elisabeth Lex}
\affiliation{%
  \institution{Graz University of Technology}
  \city{Graz} 
  \state{Austria} 
  }
\email{elisabeth.lex@tugraz.at}

\begin{abstract}
User-based Collaborative Filtering (CF) is one of the most popular approaches to create recommender systems.
This approach is based on finding the most relevant $k$ users from whose rating history we can extract items to recommend. 
CF, however, suffers from data sparsity and the cold-start problem since users often rate only a small fraction of available items. One solution is to incorporate additional information into the recommendation process such as explicit trust scores that are assigned by users to others or implicit trust relationships that result from social connections between users. Such relationships typically form a very sparse trust network, which can be utilized to generate recommendations for users based on people they trust. In our work, we explore the use of a measure from network science, i.e. regular equivalence, applied to a trust network to generate a similarity matrix that is used to select the $k$-nearest neighbors for recommending items. We evaluate our approach on Epinions and we find that we can outperform related methods for tackling cold-start users in terms of recommendation accuracy.

\end{abstract}


\fancyhead{}
\settopmatter{printacmref=false}
\maketitle

\section{Introduction}
\label{sec:intro}

Ever since their introduction, user-based Collaborative Filtering (CF) approaches have been one of the most widely adopted and studied algorithms in the recommender systems literature. CF is based on the intuition that those users, who have shown similar item rating behavior in the past, will likely give similar ratings to items in the future. %
Typically, CF comprises of three steps: first, we retrieve the $k$-nearest neighbors to the target user for whom the recommendations are generated. Second, we employ the ratings from these $k$ neighbors to determine items, which were rated highly by them but have not yet been rated by the target user. Third, these items are weighted or ranked by applying an appropriate algorithm. 

In practice, each user's ratings are stored in a rating vector. These rating vectors are then used to calculate the correlation between the target user's vector and rating vectors of the rest of the users. The higher the correlation between the rating vectors of two users, the higher their similarity. This can be assessed, e.g. via the Pearson's correlation coefficient, Cosine similarity, Jaccard index or Mean Squared Difference (MSD) \cite{sarwar2001item,liu2014new}.

However, such an approach to neighbor selection suffers from a cold-start user problem. This term refers to novel users which have rated a small number of items or have not yet rated any items at all \cite{lam2008addressing,schein2002methods}. This means that we cannot use their rating vector for finding similar users based on the pairwise vector correlation measure.

Apart from popularity-based or location-based approaches \cite{lacic2015tackling, park2009pairwise, lacic2015utilizing}, trust-based CF methods have been suggested to mitigate cold-start user problems. Their basis are trust statements expressed on platforms such as e.g. Epinions \cite{massa2007trust}. Trust statements can either be expressed explicitly via e.g. assigning trust scores or implicitly by engaging in social connections with trusted users. Based on such trust statements, trust networks can be created with the aim to generate recommendations for users based on people they trust \cite{lathia2008trust}. Since trust networks are often also sparse, a particular property of trust, namely transitivity \cite{fazeli2014}, can be exploited to propagate trust in the network. In this way, new connections are established between users, who are not directly connected, but who are connected via intermediary users.

\vspace{2mm} \noindent \textbf{The present work.} In this work, we focus on the first step of CF, i.e. finding the $k$-nearest neighbors. For this purpose, we explore the use of a similarity measure from network science referred to  "Katz similarity" (KS) by the author of \cite{Newman:2010:NI:1809753}. Although Katz himself never discussed it, KS captures regular equivalence of nodes in a network and can be applied in many different settings \cite{hasani2018consensus,helic2014regular}. As such, in this work we explore how to use KS in a trust-based CF approach.

Firstly, we utilize the trust connections to create an adjacency matrix where each entry represents a directed trust link between two users.
Secondly, we apply the KS measure on the created trust adjacency matrix. More specifically, we calculate the pairwise similarities between users by using the iterative approach on calculating KS. The iterative approach does not only allow us to calculate similarity between two nodes in the network, but additionally gives us the ability to choose the maximum used path length in doing so. This approach effectively gives us the ability to decide how far do we want to propagate trust in the network. Lastly, we use the resulting similarity matrix and apply various normalization techniques in order to get a better distribution of similarity values and better evaluation results in return. We evaluate these approaches on the Epinions dataset.

\vspace{2mm} \noindent\textbf{Contributions and findings.}
The contributions of this work are three-fold: (i) we explore the application of KS measure in the neighbor selection step of the trust-based CF approach for cold-start users, (ii) we evaluate different normalization techniques on the resulting similarity matrix to achieve better recommendation accuracy, and  (iii) we introduce an adapted KS measure that gives higher similarity values to node pairs with path lengths of $2$. In the trust-based CF setting, this means that propagated trust connections are given a higher importance than by using the standard KS measure.

Taken together, this study may help researchers to get an insight on how to apply KS on trust networks in combination with different normalization techniques to address the cold-start user problem in CF-based recommender systems. Moreover, we show that our approach for boost propagated trust values and thus increasing the impact of newly created trust connections on the recommendations could improve recommendation accuracy even further.

\section{Approach}
\label{sec:approach}

Our approach utilizes Katz similarity, which is a measure of regular equivalence, i.e. a measure of the extent to which two nodes share the same neighbors but also the extent to which their neighbors are similar. As described in \cite{global-similarity}, two nodes may have few or no neighbors in common, but they may still be similar in an indirect, global way. The idea behind KS is that paths of any length are contributing to the value of similarity between two nodes in the network, with shorter paths having a stronger impact. KS can be mathematically expressed in a matrix form as follows:

\begin{align}\label{eq:1}
    \boldsymbol{\sigma} & =\sum_{k = 0}^{\infty} (\alpha \matr{A})^k = (\matr{I} - \alpha \matr{A})^{-1}
\end{align}

\noindent where $\boldsymbol{\sigma}$ represents the similarity matrix and each value $\sigma_{i,j}$ is a similarity value between nodes $i$ and $j$, $\matr{A}$ represents the adjacency matrix of the network, $\matr{I}$ is the identity matrix which is necessary to make sure that each node is similar to itself, $\alpha$ is the attenuation factor which weights the contribution of a path of length $k$. In our trust-based setting, the adjacency matrix $\matr{A}$ is asymmetric and it represents an unweighted directed trust network, in which each node corresponds to a single user and each link represents a trust statement issued by one user to another:

\begin{align}
    A_{i,j} = 
\begin{cases}
    1,& \text{if user $j$ expressed a trust statement to user $i$} \\
    0,              & \text{otherwise}
\end{cases}
\end{align}

\noindent This also makes the similarity matrix $\boldsymbol{\sigma}$ asymmetric, which means that $\sigma_{i,j}$ does not have to be equal to $\sigma_{j,i}$, which is of advantage because in this way the asymmetric property of trust is preserved. Furthermore, one important thing to note is that for (\ref{eq:1}) to converge, the attenuation factor has to satisfy the following condition: 

\begin{align}
    \alpha < \frac{1}{\lambda_{\matr{A}}}
\end{align}

\noindent where $\lambda_\matr{A}$ is the largest eigenvalue of $\matr{A}$. The largest eigenvalue for the Epinions trust network (see Section \ref{sec:meth}) is $120.54$, hence $\alpha$ needs to be less than $0.0083$ and we set it to $0.008$ throughout all of our experiments\footnote{Since we used the iterative approach to calculating KS where $k_{max}$ was set to a small integer value, we could have chosen any $\alpha \in \left(0,1\right)$.}.
Since calculating the matrix inverse is computationally expensive, we can evaluate the above summation expression starting from $k=0$ for a fixed maximum $k$ and get the following:

\begin{align*}
\boldsymbol{\sigma}^{(0)} & = 0 \\ \boldsymbol{\sigma}^{(1)} & = \matr{I} \\ \boldsymbol{\sigma}^{(2)} & = \alpha \matr{A} + \matr{I} \\ \boldsymbol{\sigma}^{(3)} & = \alpha^2 \matr{A}^2 + \alpha \matr{A} + \matr{I} \\ & \ldots  \\ \boldsymbol{\sigma}^{(k_{max}+1)} & = \sum\limits_{k = 0}^{k_{max}} (\alpha \matr{A})^k \numberthis \label{eq:4}
\end{align*}

\vspace{2mm} \noindent\textbf{Step 1: Setting $\mathbf{k_{max}}$.}  By using this approach and setting $k_{max}$ to a positive integer value, we can define how far down the network do we want to propagate similarity or in this case, trust. In the conducted experiments we used values $1$ and $2$ as $k_{max}$, which means that we either have not propagated similarities through the network at all or that we propagated them through the network using a maximum path length of $2$.

\vspace{2mm} \noindent\textbf{Step 2: Degree normalization.} As described in \cite{Newman:2010:NI:1809753}, $\boldsymbol{\sigma}$ as defined in (\ref{eq:1}), tends to give high similarity to nodes that have a high degree. In some cases this might be desirable but if we want to get rid of this bias, we could apply a degree normalization on $\boldsymbol{\sigma}$, which would give higher similarity values to pairs of nodes that, independently of their degrees, are similar, while lower values would correspond to pairs of nodes that are dissimilar. Mathematically, for a given $k_{max}$ this step can be written as follows:

\begin{align}
\boldsymbol{\sigma}^{(k_{max}+1)}_{Dnorm} & = \matr{D}^{-1}(\sum\limits_{k = 0}^{k_{max}} (\alpha \matr{A})^k) \matr{D}^{-1} 
\end{align}

\noindent where $\matr{D}$ represents a degree matrix of a network. In the conducted experiments, we evaluated approaches with an in-degree normalization, a combined-degree normalization and without a degree normalization\footnote{Combined-degree matrix is a diagonal matrix where each value on the diagonal corresponds to the sum of in-degree and out-degree for a particular node.}.

\vspace{2mm} \noindent\textbf{Step 3: Row normalization.} After applying degree normalization, we found that all of the values in the degree normalized similarity matrix are very close to $0$, including the maximum value. Therefore, we introduced an additional step where we individually scale rows of the final resulting matrix with one of the three vector norms: $l1$, $l2$ or $max$.

\vspace{2mm} \noindent\textbf{Step 4: Boosting propagated similarities.} As already mentioned, the attenuation factor $\alpha$ is used to decrease similarity the further it gets propagated in the network. Since we set the $\alpha$ to $0.008$, similarity decays fast with each propagation step. Therefore, propagated similarity values become much smaller already in the first propagation step, i.e., for $k=2$. This would mean that trust connections created through propagation in comparison with direct trust connections have an almost insignificant impact on the resulting recommendations. Largest value for $k_{max}$ in the conducted experiments was set to $2$. This could be interpreted as using user's neighbors and their neighbors for generating item recommendations. One of the contributions of this paper was to increase the impact of propagated trust values generated with KS for $k_{max}=2$. Our proposed approach for doing so consists of the following four steps: (i) calculate $\boldsymbol{\sigma}^{(3)}$ as described above using trust network as $\matr{A}$, (ii) create a new similarity matrix $\boldsymbol{\hat{\sigma}}$ such that:

\begin{align}
    \hat{\sigma}_{i,j} = 
\begin{cases}
    \sigma^{(3)}_{i,j},& \text{if } A_{i,j} =  0\\
    0,              & \text{otherwise}
\end{cases}
\end{align}

\noindent (iii) create $\hat{\boldsymbol{\sigma}}_{norm}$ matrix by individually scaling rows of $\hat{\boldsymbol{\sigma}}$ using $l1$, $l2$ or $max$ vector norm and lastly, (iv) create a similarity matrix $\boldsymbol{\sigma}_{boost}$ such that:

\begin{align}
\boldsymbol{\sigma}_{boost} & = \matr{A} + \hat{\boldsymbol{\sigma}}_{norm}
\end{align}

\noindent With this approach, we achieve that each entry in $\boldsymbol{\sigma}_{boost}$ has a similarity value of $1$ between pairs of nodes for which there exists an explicit trust connection in $\matr{A}$ and for pairs of nodes for which the similarity has been calculated through propagation, the similarity values are not exclusively small values close to zero increasing their impact on the resulting recommendations.

\vspace{2mm} \noindent\textbf{Recommendation strategy.} As already outlined in Section \ref{sec:intro}, in this work, we focus on user-based CF.
We first create a similarity matrix using the above mentioned four steps: (i) calculate $\boldsymbol{\sigma}$ using equation (\ref{eq:4}) with $k_{max} \in \left\{ 1,2 \right\}$, (ii) normalize the similarity matrix using in-degree or combined-degree normalization, (iii) normalize similarity matrix rows using $l1$, $l2$ or $max$ vector norm, and (iv) apply boosting of propagated similarities. Steps (ii), (iii) and (iv) are optional and can be skipped. Utilizing the created trust-based similarity matrix, we first find the $k$-nearest similar users and afterwards recommend the items of those users as a ranked list of top-$N$ items to the target user $u_t$. According to the literature, the maximum number of nearest neighbors should be a value between $20$ and $60$ \cite{herlocker2002empirical}, we used $60$ in all of our experiments. The final ranking of the items to recommend is calculated as follows:

\begin{align}
pred(u_t,i) = \sum_{v \in neighbors(u_t)} sim(u_t,v)
\end{align}

\noindent where the similarity $sim(u_t,v)$ corresponds to a value from the similarity matrix calculated as proposed above.

\section{Experimental Setup} \label{sec:meth}
This section describes the experimental setup of our study including the dataset, the baseline approaches as well as the evaluation method and metrics.

\vspace{2mm} \noindent\textbf{Dataset.}
To evaluate the performance of our trust-based CF approaches for cold-start users, the well-known \textit{Epinions} dataset has been used \cite{massa2007trust}. This dataset was crawled from the consumer reviewing platform \url{Epinions.com}. Here, registered users can rate items available on the Epinions platform on a scale of $1 - 5$. Additionally, users can issue trust statements to other users on the platform, i.e., they can express how much they trust other users. In this dataset, there are only positive values for trust statements, meaning there are no negative trust statements (i.e., distrust).

Taken together, there is a total number of $49,290$ users in our dataset, which rated $139,738$ different items with $664,824$ ratings. Moreover, users have issued a total number of $487,181$ trust connections. We utilized the trust connections issued by the users to create an unweighted trust network, in which each node represents a user and each directed link represents a trust statement expressed by one user to another. The resulting trust network provides a graph density value of $0.0002$, making the trust network adjacency matrix very sparse.

\vspace{2mm} \noindent\textbf{Baseline algorithms.}
We compare our proposed approach to three baselines algorithms from the literature, which were shown to be useful methods in cold-start settings:

\textbf{$MP$.} MostPopular is a classic approach in recommender systems, which recommends the most frequently used items in the dataset to every user. Thus, it can be also applied in a cold-start setting.

\textbf{$Trust_{exp}$.} This naive trust-based approach uses explicit trust values in order to create the neighborhood of a user. Basically, adjacency matrix $\matr{A}$ created from a trust network is used as a similarity matrix which does not allow for ranking of similar users because similarity values are binary, i.e. either $0$ or $1$.

\textbf{$Trust_{jac}$.} This is a trust-based approach using Jaccard coefficient on explicit trust values and was also used by the authors of \cite{chia2011exploring}. The idea behind this approach is that two users are more similar the more trusted users they have in common. Jaccard coefficient is a statistic used to measure the similarity and diversity of sample sets and it can be written as:

\begin{align}
J(\matr{A}_{*,a}, \matr{A}_{*,b}) = \frac{|\matr{A}_{*,a} \cap \matr{A}_{*,b}|}{|\matr{A}_{*,a} \cup \matr{A}_{*,b}|}
\end{align}

\noindent where $J(\matr{A}_{*,a}, \matr{A}_{*,b})$ is used to calculate similarity between users $a$ and $b$, $\matr{A}_{*,a}$ corresponds to explicit values given to other users in the trust network by user $a$ and the same applies to $\matr{A}_{*,b}$ for user $b$.

\vspace{2mm} \noindent\textbf{Evaluation method and metrics.}
In order to compare our proposed approach to these baseline algorithms in a cold-start setting, we extracted all users with no more than 10 rated items from the dataset. This resulted in $25,393$ users, for which we put all of their rated items into the test set. To finally quantify the performance of our evaluated algorithms, we used the well-established accuracy metrics $nDCG$, $Precision$ and $Recall$ for $k = 1 - 10$ recommended items \cite{SmythMcClave01,herlocker2004evaluating}.

\begin{table}[t]
\setlength{\tabcolsep}{5pt}
\renewcommand{\arraystretch}{1.3}

\centering
\scalebox{.8}{
\begin{tabular}{|c|c|c|c|c||c|c|c|}
\hline
\multirow{2}{*}{Algorithm} & 
\multirow{2}{*}{$k_{max}$} & 
Degree & 
Row & 
\multirow{2}{*}{Boost} & 
\multirow{2}{*}{nDCG} & 
\multirow{2}{*}{R} & 
\multirow{2}{*}{P} \\ 
&& norm. & norm. &&&&\\ \hline \hline

\multicolumn{5}{|c||}{$Trust_{exp}$} & .0224                      & .0296                       & .0110                                               \\ \hline
\multicolumn{5}{|c||}{$Trust_{jac}$} & .0176                      & .0219                       & .0087                                               \\ \hline 
\multicolumn{5}{|c||}{$MP$} & .0134                      & .0202                       & .0070                                        \\ \hline \hline

$KS_{PCMB}$ & 2  & Combined      & Max                     & Yes     & .0303                     & .0425                       & .0117                                                 \\ \hline  
$KS_{PCMN}$ & 2 & Combined      & Max                      & No      & .0295                     & .0422                       & .0113                                                 \\ \hline
$KS_{PCL_1B}$ & 2 & Combined      & L1                       & Yes     & .0273                     & .0358                       & .0106                                                 \\ \hline
$KS_{PNL_2B}$ & 2 & No degree     & L2                       & Yes     & .0257                     & .0340                       & .0106                                                 \\ \hline
$KS_{NCMN}$ & 1 & Combined      & Max                      & No      & .0213                     & .0289                       & .0106                                                 \\ \hline
$KS_{NINN}$ & 1 & In degree     & N/A                       & No      & .0161                     & .0243                       & .0087                                                 \\ \hline
$KS_{PNNN}$ & 2 & No degree     & N/A                      & No      & .0036                     & .0057                       & .0020                                              \\ \hline
\end{tabular}
}
\caption{Evaluation results for $k=10$. The first three rows refer to the baseline approaches while the remaining ones correspond to our KS-based approaches. The best performing approach in terms of all accuracy measures was $KS_{PCMB}$, where we used combined degree normalization, row normalization with $max$ norm as well as trust propagation (i.e., $k_{max}=2$) with boosting of the propagated similarity values.
\vspace{-7mm}
}

\label{results-table}
\end{table}
\renewcommand{\arraystretch}{1}

\section{Results}

In our study, we evaluated $33$ approaches for all possible step combinations when creating the similarity matrix (see Section \ref{sec:approach}). However, for the sake of space, in Table \ref{results-table}, we only report the results for a subset of these approaches that provide the most insightful findings. All of the evaluation results are reported for $k=10$, i.e. for $10$ recommended items. As it can be seen in Table \ref{results-table}, the best performing approach in terms of all accuracy measures was $KS_{PCMB}$, where we used a combined degree normalization, row normalization with $max$ norm as well as trust propagation ($k_{max}=2$) with boosting of the propagated similarity values.

One of the most interesting findings was that if similarity propagation was not used, i.e., $k_{max}$ was set to $1$, better results were achieved if no degree normalization was applied as well as no row normalization. This means that if similarity propagation was not applied, it was better to simply use the $Trust_{exp}$ baseline approach. However, if $k_{max}$ was set to $2$, we noticed result improvements in almost all of the cases except when no row normalization was applied, e.g., in the case of $KS_{PNNN}$.

Additionally, similarity propagation with $k_{max}=2$ increased the similarity matrix density from $0.0002$ to $0.008$. It turned out that row normalization was a very important step in using KS with similarity propagation for neighbor selection. Another important finding was that the combined-degree normalization provided better results than in-degree normalization in most of the cases. Also, with respect to row normalization, $max$ norm provided better results than $l1$ and $l2$ in most of the cases. Finally, with degree normalization and row normalization unchanged, boosting of propagated similarities often provided better results.

\begin{figure}[t]
\centering
\includegraphics[width=.48\textwidth]{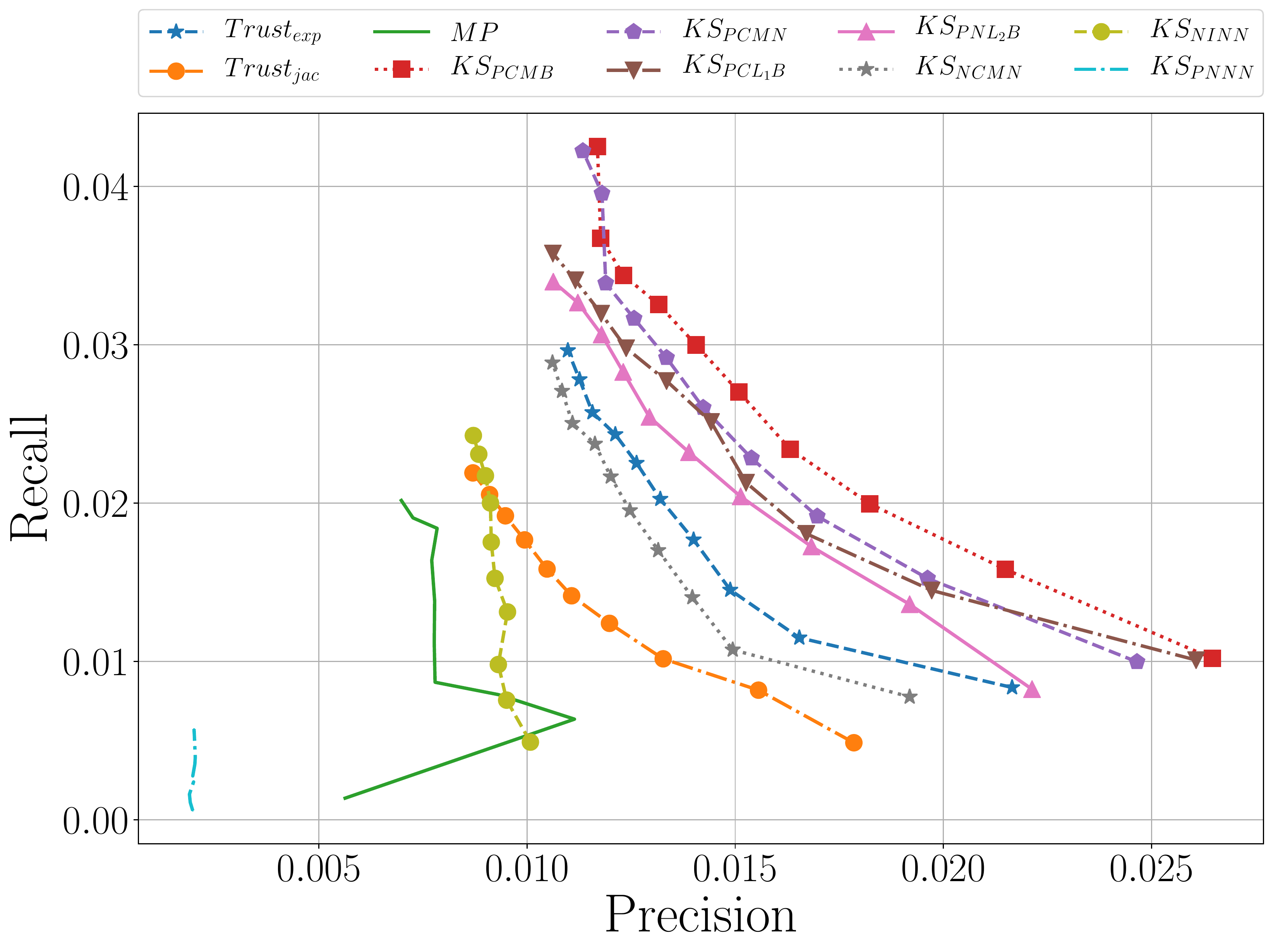}
\caption{Recommender accuracy of the described approaches in the form of Recall-Precision plots for $k=1-10$ recommended items. Again, we can observe that our $KS_{PCMB}$ approach outperforms all three baselines as well as the other KS-based algorithms.
\vspace{-3mm}
}
\label{fig:recall-precision}
\end{figure}

Finally, in Figure \ref{fig:recall-precision}, we show the performance of all approaches listed in Table \ref{results-table} in form of Recall-Precision plots for different number of recommended items (i.e., $k = 1 - 10$). The results clearly show that the best performing algorithm (i.e., $KS_{PCMB}$) again outperforms all of the other approaches also for a smaller number of recommended items (i.e., for $k < 10$). Precisely, the same conclusion can be drawn for all KS-based approaches which already outperformed the baselines in Table \ref{results-table}, thus consistently providing more accurate recommendations than the baselines.

\newpage
\section{Conclusion \& Future Work}
\label{sec:conc}

In this paper, we explored the use of Katz similarity (KS), a similarity measure of regular equivalence in networks, for selecting $k$-nearest neighbors in a Collaborative Filtering (CF) algorithm for cold-start users. We used an iterative approach for calculating KS since it provides the ability to restrict the length of paths in the network used for similarity calculation. We found that KS can be a very useful measure for neighbor selection if it is used with degree-normalization and row normalization, especially when using similarity propagation. When these techniques are properly combined with KS, we managed to outperform related approaches for tackling the cold-start problem. Our results also indicate that trust propagation is a very important feature when using trust networks in a CF setting as well as that KS is a useful technique for efficiently propagating trust in a network. Summed up, our study may help researchers to get an insight on how to apply KS on trust networks in combination with different normalization techniques to address the cold-start user problem in recommender systems.

One limitation of this study was that we only evaluated our approaches using recommender accuracy, although optimizing on non-accuracy measures has been closely tied to user satisfaction \cite{zhang2012auralist,lacic2017beyond}. As such, in the future we plan to investigate the impact of trust-based networks on beyond accuracy metrics such as novelty, diversity and coverage. Moreover, we also plan to explore the use of recently popularized node embeddings (e.g., Node2Vec \cite{grover2016node2vec}) for trust networks to further improve our results.

\vspace{2mm} \noindent\textbf{Acknowledgments.} This work was supported by the Know-Center (Austrian COMET program) and the AFEL project (GA: 687916).

\newpage

\bibliographystyle{abbrv}

\end{document}